\documentclass[aps,prb,10pt,twocolumn,amsmath,amssymb,superscriptaddress,showpacs,preprintnumbers,footinbib,reprint]{revtex4-1}

\usepackage[T1]{fontenc}
\usepackage{textcomp}
\usepackage{times}
\usepackage{bm}
\usepackage{natbib}
\usepackage{graphicx,color,epsfig}
\usepackage{xspace}
\usepackage{stmaryrd}
\usepackage{dcolumn} 
\usepackage{hyperref}
\hypersetup{
  colorlinks   = true,
  citecolor    = blue,
  linkcolor    = blue,
  urlcolor     = blue
  }
\usepackage[usenames,dvipsnames,svgnames,table]{xcolor}
\DeclareFontFamily{OT1}{pzc}{}
\DeclareFontShape{OT1}{pzc}{m}{it}{<-> s * [1.2] pzcmi7t}{}
\DeclareMathAlphabet{\mathpzc}{OT1}{pzc}{m}{it}

\newcommand{\ham}{\mathpzc{H}}
\newcommand{\tr}{\mathrm{Tr\,}}
\newcommand{\wnsq}{\langle{w^2_B}\rangle}
\newcommand{\ket}[1]{|{#1}\rangle}
\newcommand{\bra}[1]{\langle{#1}|}

\newcommand{\grant}[1]{{\color{black} #1}}

\begin{document}
\date{\today}

\author{Yoshitomo Kamiya}
\affiliation{
  Interdisciplinary Theoretical Science Research Group (iTHES)
  and Condensed Matter Theory Laboratory, RIKEN,
  Wako, Saitama 351-0198, Japan
}
\author{Yasuyuki Kato}
\affiliation{
  RIKEN Center for Emergent Matter Science (CEMS),
  Wako, Saitama 351-0198, Japan
}
\affiliation{
  Department of Applied Physics, University of Tokyo, 
  Bunkyo, Tokyo 113-8656, Japan
} 
\author{Joji Nasu}
\affiliation{
  Department of Physics, Tokyo Institute of Technology, 
  Meguro, Tokyo 152- 8551, Japan
} 
\author{Yukitoshi Motome}
\affiliation{
  Department of Applied Physics, University of Tokyo, 
  Bunkyo, Tokyo 113-8656, Japan
} 

\title{Magnetic three states of matter: A quantum Monte Carlo study of spin liquids}
\begin{abstract}
  We present thermodynamic phase diagrams showing magnetic analog of ``three states of matter,'' namely, spin liquid, paramagnetic, and magnetically ordered phases, obtained by unbiased quantum Monte Carlo simulations. Our simulations are carried out for Kitaev's toric codes in two and three dimensions, i.e., the simplest realizations of gapped topological $Z_2$ spin liquids, extended by a nearest-neighbor ferromagnetic Ising coupling. We find that the ordered phase borders on the spin liquid by a discontinuous transition line in three dimensions, while it grows continuously from the quantum critical point in two dimensions. In both cases, our results elucidate peculiar proximity effects to the nearby spin liquids in the high-temperature paramagnetic phase, even when the ground state is magnetically ordered. The thorough study of magnetic three states of matter is achieved by introducing the ``fictitious vertex'' method into the directed loop algorithm. This provides a generic prescription to simulate models with off-diagonal multispin interactions, in which the conventional scheme may suffer from intrinsic ergodicity breakdown as in the present case.
\end{abstract}


\maketitle

\textit{Introduction.}
The concept of topological order becomes central for exploring new states of matter~\cite{Wen2004}. Particularly, the possibility of quantum spin liquids (QSLs) with topological orders has been extensively discussed to explain experiments in highly frustrated quantum magnets in which conventional symmetry-breaking order seems prevented much below the Curie-Weiss temperature~\cite{Balents2010}.
Naturally, more theoretical inputs for verifying QSLs are desired. Whether or not QSLs have unmistakable hallmark is one of the most relevant questions. Also, even if a system under investigation may develop a magnetic order at the lowest temperature, we may ask if there are any characteristic proximity effects to a nearby possible QSL phase. To answer these questions,
the Kitaev model~\cite{Kitaev2003,Kitaev2006} and its descendants are useful as their ground states are
QSLs, providing good starting points for exploring quantum and thermal effects~\cite{Trebst2007,Tupitsyn2010,Dusuel2011,Wu2012,Quinn2015,Jackeli2009,Chaloupka2010,Chaloupka2013,Kimchi2014a,Kimchi2014b,Sela2014,Reuther2011,Yamaji2014}.
Some of them may have connections with real materials such as iridates after including perturbations stabilizing magnetic orders~\cite{Jackeli2009,Chaloupka2010,Chaloupka2013,Kimchi2014a,Kimchi2014b,Sela2014,Reuther2011,Yamaji2014}. Recently, two of the authors and their collaborator investigated thermal effects in the Kitaev models on honeycomb~\cite{Nasu2015a}, hyperhoneycomb~\cite{Nasu2014a,Nasu2014b}, and decorated honeycomb~\cite{Nasu2015b} lattices, clarifying the nature of QSL-paramagnetic transitions, which is very sensitive to types of QSLs and dimensionality.

A combination of thermal and quantum effects on Kitaev-type models is a more challenging topic especially in three dimensions (3D). To the best of our knowledge, effects of a magnetic field were investigated at both zero and finite temperature ($T$) in two dimensions (2D)~\cite{Trebst2007,Tupitsyn2010,Dusuel2011,Wu2012}, but not in 3D. As for \textit{exchange}-type perturbations, although they pose interesting questions regarding transitions between topological and conventional ordered phases, the corresponding study at finite $T$ was not initiated until a recent functional renormalization-group calculation in 2D~\cite{Reuther2011}. The studies in 3D were so far restricted to semiclassical treatments~\cite{EKHLee2014,SBLee2014} or the Bethe lattice~\cite{Kimchi2014b}.

In this Rapid Communication, we wish to initiate a full quantum treatment of the exchange-type interaction effects on a 3D QSL at finite $T$. We also investigate a 2D model and compare the phase diagrams. We will focus on gapped $Z_2$ QSLs, which may be realized in certain frustrated Heisenberg systems (e.g., 2D kagome~\cite{Depenbrock2012} and 3D hyperkagome~\cite{Lawler2008}) or iridates (e.g., related compounds with Li$_2$IrO$_3$~\cite{Nasu2014a,Modic2014,Takayama2015}). Since our interest is in generic properties of QSLs, we will consider Kitaev's toric codes on square~\cite{Kitaev2003} and cubic~\cite{Hamma2005,Castelnovo2008,Nussinov2008} lattices as the simplest Hamiltonians realizing the gapped $Z_2$ QSLs. We add a nearest-neighbor ferromagnetic (FM) Ising coupling, which introduces quantum fluctuations to otherwise static flux excitations of the QSL states. These quasiparticles are known to have nontrivial properties, such as anyonic mutual statistics in 2D (Ref.~\onlinecite{Kitaev2003}) or a confinement-deconfinement transition in 3D~\cite{Castelnovo2008,Nussinov2008,Nasu2014a,Nasu2014b}.
  
Our phase diagram obtained by unbiased quantum Monte Carlo (QMC) methods
completes a study of the magnetic analog of ``three states of matter,'' namely, QSL (``liquid''), paramagnetic (``gas''), and FM (``solid'') phases, while the first two were the subjects of Refs.~\onlinecite{Nasu2014a,Nasu2014b,Nasu2015a}.
We find a qualitative difference between 2D and 3D and elucidate the distinct nature of quasiparticles responsible for the difference. We also unveil proximity effects to the QSL phases even when the ground state is magnetically ordered.
Although conventional QMC algorithms turn out to be inefficient in and near the QSL phase, or even nonergodic with the directed loop updates~\cite{Syljuasen2001,Kawashima2004,Alet2005}, we solve this issue by introducing the ``fictitious vertex'' method, a generic idea to
handle off-diagonal multispin interactions in QMC in a simple manner.

\begin{figure}[t]
  \vspace{-4pt}
  \includegraphics[width=0.9\hsize]{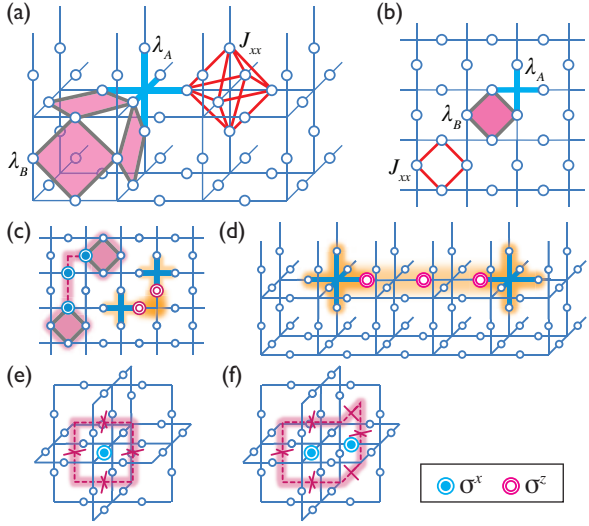}
  \caption{\label{fig:model} (Color online)
    (a) 3D and (b) 2D lattices for $\ham$ in Eq.~\eqref{eq:H} with circles representing $S=1/2$ spins.
    (c) Electric charge $A_s = -1$ (crosses) and magnetic flux $B_p = -1$ (plaquettes) excitations in 2D at the ends of strings of $\sigma^{z}$ and $\sigma^{x}$ operators, respectively.
    (d) Electric charge excitations in 3D.
    (e) and (f) Examples of magnetic flux loops in 3D
    with crosses indicating plaquettes with $B_p = -1$.
  }
  \vspace{-9pt}
\end{figure}

\textit{Model.}
Both in 2D and 3D, the Hamiltonian can be formally written as
\begin{align}
  \ham = -\lambda_A \sum_{s} A_s - \lambda_B \sum_{p} B_p - J_{xx} \sum_{\langle{ij}\rangle} \sigma_i^x \sigma_j^x,
  \label{eq:H}
\end{align}
where $S=1/2$ spins [$\bm{\sigma}_i=(\sigma_i^x,\sigma_i^y,\sigma_i^z)$ are Pauli matrices]  reside on links and $\langle{ij}\rangle$ denotes nearest neighbors. $\lambda_A$, $\lambda_B$, and $J_{xx}$ are positive. $A_s = \prod_{j\in s} \sigma_j^x$ denotes a four- (six-)spin product of $\sigma^x$ assigned to a node $s$ of the square (cubic) lattice, while $B_p = \prod_{j\in p} \sigma_j^z$ is a four-spin plaquette operator [Figs.~\ref{fig:model}(a) and \ref{fig:model}(b)]. Since $[A_s,B_p]=0$ $\forall s$ and $\forall p$, both in 2D and 3D, these are local conserved quantities of $\ham_0 = -\lambda_A \sum_{s} A_s - \lambda_B \sum_{p} B_p$ with eigenvalues $\pm 1$, though not all of them are independent~\cite{Kitaev2003,Castelnovo2008}. $\ham_0$ is exactly solvable and the condition for a ground state is that all of these eigenvalues are $+1$. In addition, there are nonlocal $Z_2$ operators [global string operators in 2D (Ref.~\onlinecite{Kitaev2003}) and either global string or global surface operators in 3D (Ref.~\onlinecite{Castelnovo2008})] commuting with any $A_s$ and $B_p$, and thus with $\ham_0$. They distinguish $2^d$-fold degeneracy of the spectrum of $\ham_0$ on the $d$-dimensional torus~\cite{Kitaev2003,Castelnovo2008}. The FM Ising coupling $J_{xx}$ makes $B_p$'s unconserved (see below). For small $J_{xx}$, splitting of the topological ground-state degeneracy is exponentially small in the system size $L$ (Ref.~\onlinecite{Kitaev2003}).

In both dimensions, ``electric charges,'' or nodes with $A_s = -1$, can be created or annihilated in pairs by $\sigma^z_i$ ($i \in s$) and they are deconfined quasiparticles [Figs.~\ref{fig:model}(c) and \ref{fig:model}(d)]. A crucial difference between 2D and 3D appears in ``magnetic fluxes'' corresponding to plaquettes with $B_p = -1$, created or annihilated by $\sigma^x_i$ with $i \in p$
[Figs.~\ref{fig:model}(c), \ref{fig:model}(e), and \ref{fig:model}(f)].
While fluxes in 2D are pointlike and deconfined, they always form \textit{loops} in 3D as the operator identity $\prod_{p \in \text{cube}} B_p = 1$ implies that a flux string entering an elementary cube must leave it (no magnetic monopole). To stretch a flux loop, additional plaquettes must be flipped with an extra energy linear in the increased length [Fig.~\ref{fig:model}(e)]. Consequently, loops are confined at low $T$. More delocalized loops appear at higher $T$ favored by entropic effects, suggesting that a proliferation of flux loops may occur at a finite $T$.
In fact, it was shown for $J_{xx} = 0$ that this corresponds to a finite-$T$ transition~\cite{Castelnovo2008,Nussinov2008}. Topological entanglement entropy is finite for $T < T_{c}$~\cite{Castelnovo2008}, although expectation values of the nonlocal $Z_2$ topological order parameters vanish for any finite $T$~\cite{Castelnovo2008,Nussinov2008}.

\begin{figure}[t]
  \includegraphics[width=0.975\hsize]{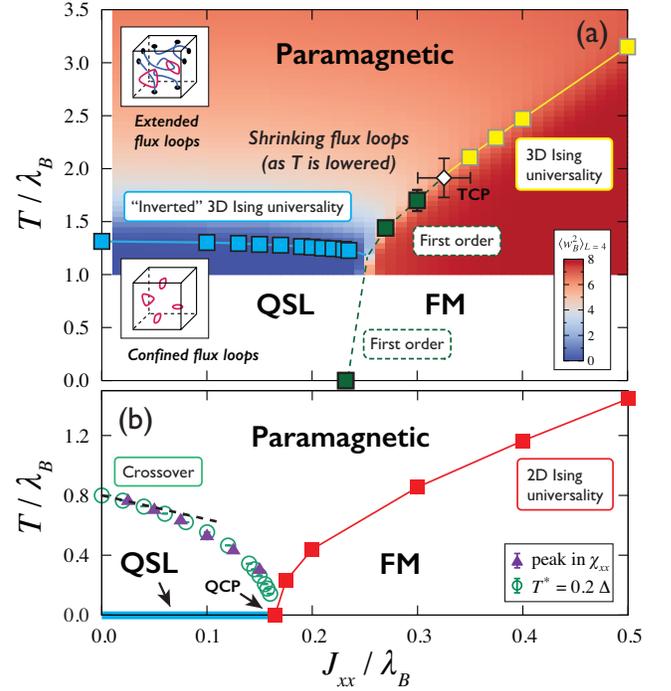}
  \vspace{-2pt}
  \caption{\label{fig:phase diagrams} (Color online)
    (a) Phase diagram in 3D for $\lambda_A = \lambda_B$. Some error bars are smaller than the symbol size. The lines are guides to the eye. The intensity plot, presented for $T / \lambda_B > 1$, corresponds to $\langle{w_B^2}\rangle$ for $L = 4$ [see also Fig.~\ref{fig:results1}(c)].
    (b) Phase diagram in 2D for $\lambda_A = \lambda_B$. The dashed line shows the asymptotic behavior of $T^\ast \equiv 0.2\Delta \sim 0.8 \lambda_B - 1.6 J_{xx}$, where the prefactor ($=0.2$) is chosen to match $T^\ast$ with the peak position of $\chi_{xx}$.
  }
  \vspace{-8pt}
\end{figure}

This transition is expected to be of the inverted 3D Ising universality class. This can be explained by the duality between the 3D classical Ising model on the simple cubic lattice and the flux sector of the 3D toric code. By using the high-$T$ expansion, the partition function of the Ising model can be rewritten as a sum over loop configurations: $Z_\text{Ising} \propto \sum_\text{loops} (\tanh K)^{\ell_\text{total}}$ where $K$ and $\ell_\text{total}$ are the dimensionless coupling and the total loop length of a configuration, respectively. Thus, these models can be related through $\exp(-2\lambda_B / T) = \tanh K$, with a minor caveat that the winding number of flux loops (high-$T$ graphs) is even (can be even or odd), which however would not alter the universality class~\cite{Nasu2014a}. Consequently, the two models should undergo a phase transition of the same (i.e., 3D Ising) universality class. However, this is with the \textit{inverted} $T$ axes in a sense that confined and extended loops are favored, respectively, in the low- and high-$T$ (high- and low-$T$) phases in the 3D toric code (Ising model).
Hence, universal amplitude ratios of one model are inverses of the other.

\textit{Phase diagram in 3D.}
For $J_{xx} \ne 0$, the flux loops acquire quantum dynamics, while electric charges remain static. The model is no longer exactly solvable, and we perform continuous time world-line QMC simulations. We show the obtained phase diagram in 3D in Fig.~\ref{fig:phase diagrams}(a), leaving the details of our method for later. Below, we will discuss the case of $\lambda_A = \lambda_B$, where fluxes and electric charges have comparable excitation energies for small $J_{xx}$.

\begin{figure}[t]
  \includegraphics[width=0.975\hsize]{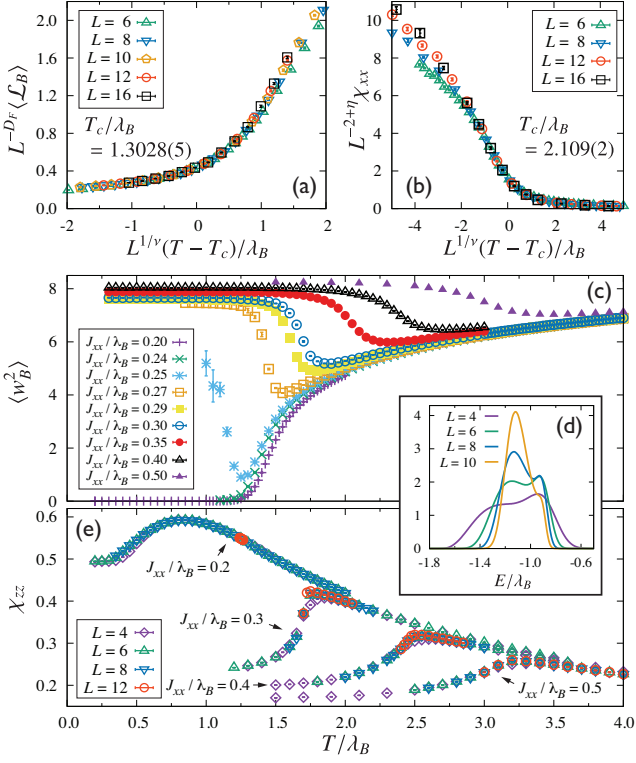}
  \caption{\label{fig:results1}(Color online)
    FSS plots of (a) $\langle{\mathcal{L}_B}\rangle$ for $J_{xx} / \lambda_B = 0.1$ and (b) $\chi_{xx}$ for $J_{xx} / \lambda_B = 0.35$, respectively, produced by assuming $D_F = 1.7349(65)$~\cite{Winter2008}, $\eta = 0.036\,39(15)$, and $\nu = 0.630\,12(16)$~\cite{Campostrini2002}.
    (c) $\wnsq$ for $L = 4$.
    (d) The histogram of the energy density for $J_{xx}/\lambda_B = 0.3$ at $T / \lambda_B = 1.7$.
    (e) $\chi_{zz}$ for $J_{xx}/\lambda_B = 0.2$, $0.3$, $0.4$, and $0.5$.
    All the data correspond to the 3D model with $\lambda_A = \lambda_B$.
  }
  \vspace{-6pt}
\end{figure}

The paramagnetic-QSL transition line cannot be characterized by any local order parameter. Instead, we study the flux loop proliferation by evaluating the expectation value of length $\langle{\mathcal{L}_B}\rangle$ of the largest loop. We confirm the inverted 3D Ising universality by finite-size scaling (FSS) analysis of $\langle{\mathcal{L}_B}\rangle$, which leads to excellent data collapse [Fig.~\ref{fig:results1}(a)].
Here, we estimate $T_c$ by using the Bayesian FSS method~\cite{Harada2011,Harada2015} assuming $\nu = 0.630\,12(16)$~\cite{Campostrini2002} and the fractal dimension $D_F = 1.7349(65)$ estimated for the high-$T$ graphs in the 3D Ising model~\cite{Winter2008}. We find that $T_c$ decreases only slightly with $J_{xx}$, suggesting that the effective flux-loop tension is weakly renormalized by $J_{xx}$.
As a different measure of delocalization of flux loops, we calculate $\wnsq$ defined by the summation over the squared winding number of each flux loop $\mathcal{C}$, i.e., $w^2_B = \sum_\mathcal{C}\sum_{\mu\in\{x,y,z\}} [L^{-1}\oint_{\,\mathcal{C}} \mathbf{a_\mu} \cdot d\mathbf{s}]^2$, where $\mathbf{a_\mu}$ denotes the unit vector along the $\mu$ axis.
$\wnsq$ is expected to be zero (nonzero) for $T < T_c$ ($T > T_c$) in the thermodynamic limit~\cite{Nasu2014a}. In the QSL regime ($J_{xx} / \lambda_B \lesssim 0.24$), we find $\wnsq$ decreasing continuously and monotonically as
lowering $T$
[Fig.~\ref{fig:results1}(c)], suggesting that the paramagnetic-QSL transition is always of second order in this model.

The FM phase with $\langle{\sigma^x}\rangle \ne 0$ is stabilized for large $J_{xx}$. For sufficiently large $J_{xx}$, the finite-$T$ transition is in the 3D Ising universality class as confirmed by FSS analysis of the $x$-component magnetic susceptibility $\chi_{xx}$ [Fig.~\ref{fig:results1}(b)]. The transition changes from second to first order for small $J_{xx}$, as indicated by the double peaks in the energy histogram in Fig.~\ref{fig:results1}(d). This suggests that there is a tricritical point $(T^\text{TCP}, J_{xx}^\text{TCP})$ on the paramagnetic-FM phase boundary [Fig.~\ref{fig:phase diagrams}(a)].

On the FM side of the phase diagram, $\wnsq$ shows a dip enhanced near the QSL-FM phase boundary [Fig.~\ref{fig:results1}(c)]. Here, $\langle w^2_B \rangle$ has a sizable magnitude inside the FM phase as in the paramagnetic phase, simply because $\langle{\sigma^x}\rangle \ne 0$ implies a random configuration of $\sigma_i^z$. For $J_{xx} > J_{xx}^\text{TCP}$, the upturn of the broad dip is associated with the onset of the critical phenomena of the FM transition. For $J_{xx} < J_{xx}^\text{TCP}$, the upturn of the pronounced dip is suggested to become a jump in the thermodynamic limit.
By comparing with the behavior on the QSL side, we conclude that the salient dip in $\langle w^2_B \rangle$, implying
flux loops shrinking
with decreasing $T$ above $T_c$, is a proximity effect to the QSL phase.
The proximity effect to the QSL phase can also be observed in the $z$-component magnetic susceptibility $\chi_{zz}$. As shown in Fig.~\ref{fig:results1}(e), $\chi_{zz}$ develops a more pronounced peak above $T = T_c$ with decreasing $J_{xx}$ on the FM side.

\begin{figure}[t]
  \includegraphics[width=0.975\hsize]{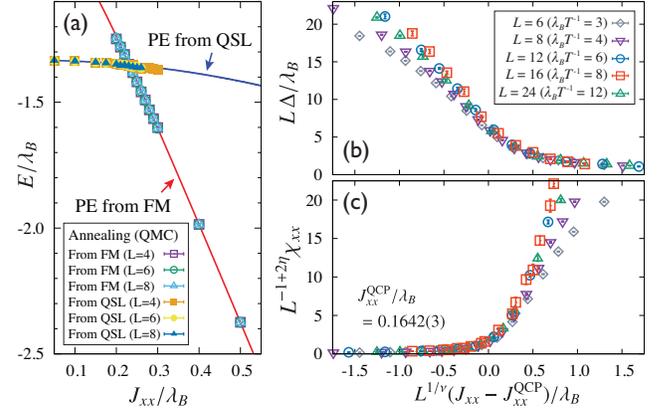}
  \caption{\label{fig:results2}(Color online)
    (a) Estimated ground-state energy in 3D, compared with the results of the second-order perturbative expansions (PEs).
    (b) FSS plot of the energy gap $\Delta$ and (c) $\chi_{xx}$ to analyze quantum critical behavior in 2D assuming $\eta = 0.03639(15)$, $\nu = 0.63012(16)$~\cite{Campostrini2002}, and $z = 1$.
  }
  \vspace{-6pt}
\end{figure}

The QSL-FM quantum phase transition is strongly first order in this model. Figure~\ref{fig:results2}(a) shows the evolution of energy density at $T/\lambda_B = 4 / L$ ($T \to 0$ as $L \to \infty$) from both sides of the phase diagram, revealing a level crossing at $J_{xx}^\text{cross} / \lambda_B = 0.232(2)$. By comparing the data with second-order perturbation theory, $E_\text{small-$J_{xx}$} = -\lambda_A/3 - \lambda_B - J_{xx}^2 / (3\lambda_B)$ and $E_\text{large-$J_{xx}$} = -\lambda_A/3 - 4J_{xx} - \lambda_B^2 / (48 J_{xx})$ for small and large $J_{xx}$, respectively, we find a good agreement indicating that both phases are rather stable up to $J_{xx} = J_{xx}^\text{cross}$. 

\textit{Phase diagram in 2D.}
The fluxes and electric charges in 2D are both pointlike (no loop structure) and deconfined~\cite{Kitaev2003} [Fig.~\ref{fig:model}(c)]. Consequently, infinitesimal thermal fluctuations can destroy the QSL state. At $T=0$, the QSL is stable against local perturbations because these excitations are gapped. We evaluate the gap $\Delta$ by applying the second-moment method~\cite{Cooper1982} to $\langle{T_\tau \sigma^x_{\mathbf{k}=0}(\tau) \sigma^x_{\mathbf{k}=0}(0)}\rangle$ as this corresponds to a propagator of a flux pair for small $J_{xx}$. This quantity should be associated with the lowest-energy gap also near the QSL-FM quantum transition, as $\langle{\sigma^x}\rangle$ becomes nonzero at this point. Our FSS analysis of $\Delta$ suggests a quantum critical point (QCP) in the (2+1)D Ising universality class at $J_{xx}^\text{QCP} / \lambda_B = 0.1642(3)$ [Fig.~\ref{fig:results2}(b)].
Near the QCP, our results strongly suggest $\langle{A_s}\rangle=1$ as $T \to 0$ and $L \to \infty$, providing numerical evidence that an effective description of the QCP is that of the (2+1)D $Z_2$ gauge theory~\cite{Kogut1979}, consistent with the gapped spectrum in the electric charge sector. This observation motivates us to introduce dual Pauli matrices~\cite{Kogut1979} relating $B_p = \mu_p^z$ and $\sigma^x_{i_{pp'}} = \mu_p^x \mu_{p'}^x$ ($i_{pp'}$ denotes a site between neighboring plaquettes $p, p'$), and we obtain two copies of the transverse-field Ising models. This duality argument supports our observation of the (2+1)D Ising universality.

The mapping implies that the primary order parameter at the QCP is not $\langle{\sigma^x_{i}}\rangle$ but its \textit{fractionalized} component $\langle{\mu_p^x}\rangle$ (while the order parameter at finite-$T$ transition is simply $\langle{\sigma^x_{i}}\rangle$).
Since the two $\mu^x$ operators comprising $\sigma^x$ belong to the decoupled copies in the dual description, the scaling dimension of $\langle{\sigma^x_{i}}\rangle$ is doubled compared to that of the primary field; thus we conclude that the magnetic susceptibility does not follow the standard scaling $\chi_{xx} \sim L^{2 - \eta}$ but $\chi_{xx} \sim L^{1 - 2\eta}$, as we confirm in Fig.~\ref{fig:results2}(c). Such unconventional scaling can be seen as another nontrivial proximity effect to QSL in 2D, and should be observable as Curie-Weiss-like scaling in the quantum critical regime at finite $T$: $\chi_{xx} \sim T^{-1 + 2\eta}$. Here we note that other quantities, such as the specific heat, exhibit clear deviations from the paramagnetic behavior. 
We also evaluate $\chi_{xx}$ on the QSL side, where we find a non-divergent peak at $T \approx 0.2\Delta$.
This defines crossover temperature $T^\ast$ [Fig.~\ref{fig:phase diagrams}(b)], below which thermodynamic properties are very similar to the ground-state properties. The FM phase is stable at $T>0$, as expected for a system with global $Z_2$ symmetry.

\textit{Methods.}
We outline the idea of \textit{fictitious vertices} that we introduced in the current study~\cite{tbp} in the framework of the directed loop algorithm (DLA)~\cite{Syljuasen2001,Kawashima2004,Alet2005}.
We used the loop algorithm~\cite{Evertz2003,Kawashima2004} in the basis diagonalizing $\sigma_i^x$'s for large $J_{xx}$ supplemented by cluster-type updates for elementary cubes in 3D.
However, the loop algorithm leads to huge clusters in the QSL regime despite the very short correlation length,
which reduces the efficiency of the simulation. While such a difficulty is usually solved by the DLA, the conventional DLA encounters an \textit{intrinsic} ergodicity breakdown in the present model due to the off-diagonal multispin vertices.
The fictitious vertex method can recover the broken ergodicity. This method is not necessarily a problem-specific idea, but a rather generic one~\cite{tbp} to handle models with off-diagonal multispin vertices, such as spin-orbital~\cite{Kugel1982} or ring-exchange terms~\cite{Sandvik2002,Melko2004,Melko2005,Dang2011,Rousseau2013}.

\begin{figure}[t]
  \includegraphics[width=0.975\hsize]{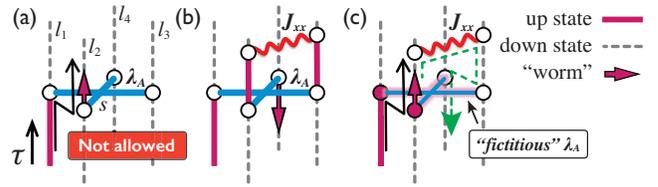}
  \caption{\label{fig:method}(Color online)
    (a) Worm-hopping at a $\lambda_A$ vertex [illustrated in (2+1)D] not allowed in the conventional DLA because $\bra{\downarrow_{l_1}\uparrow_{l_2}\downarrow_{l_3}\downarrow_{l_4}}A_s + 1\ket{\uparrow_{l_1}\downarrow_{l_2}\downarrow_{l_3}\downarrow_{l_4}} = 0$.
    (b) Example of a local configuration that cannot be generated in the conventional DLA.
    (c) By promoting the vertex into a \textit{fictitious vertex}, the hopping of the type (a) can be included;
    the configurations like (b) can be realized if the subsequent worm move coincides with the directed dashed line.
  }
  \vspace{-4pt}
\end{figure}

We work on the $\sigma^z$ basis, as this is convenient to measure fluxes. In the DLA~\cite{Syljuasen2001,Kawashima2004,Alet2005}, the world-line configurations contributing to  $g_{i_1 i_2}(\tau_1,\tau_2) \equiv \tr T_\tau \exp(-\beta \ham) \sigma^x_{i_1}(\tau_1) \sigma^x_{i_2}(\tau_2)$ are sampled by moving one of the external lines (the ``worm''), say at $(i_1,\tau_1)$, while fixing the other at $(i_2,\tau_2)$. At each space-time point, there is either the identity or one of the vertex operators, i.e., $A_s + 1$, $B_p$, or $\sigma_i^x\sigma_j^x + 1$ (which we call $\lambda_A$, $\lambda_B$, and $J_{xx}$ vertices, respectively, and the constants are absorbed into $\ham$), resulting from an expansion of $\exp(-\epsilon\ham)$ to $O(\epsilon)$ with $\epsilon \ll T^{-1}$.
Every off-diagonal event in a world-line configuration corresponds to a vertex operator and its nonzero matrix element. The worm, locally updating the configuration as it moves, may be scattered at a vertex if the resulting state corresponds to a nonzero matrix element. However, an issue here is that this conventional scheme prohibits a hopping at $\lambda_A$ vertices [Fig.~\ref{fig:method}(a)]. Consequently, it is impossible to generate off-diagonal processes that can arise from combinations of $\lambda_A$ and $J_{xx}$ vertices [Fig.~\ref{fig:method}(b)]. This causes an intrinsic ergodicity issue in the conventional algorithm.

Fictitious vertices allow such otherwise prohibited hopping by extending the configuration space. Similar ideas were discussed in Refs.~\onlinecite{Kato2013,Soyler2009}. For example, if the worm (at site $l$) reaches a $\lambda_A$ vertex (at $s$), we promote this into a fictitious vertex $(A_s + 1)\sigma_{l}^x \sigma_{l'}^x$ with a probability $1-p$, where $l'\,(\ne l) \in s$ is chosen randomly (with a probability $p$, the worm follows the conventional scheme). With the aid of the attached $\sigma^x$ operators marking $l$ and $l'$, the worm hops from $l$ to $l'$ and leaves the vertex in $\pm \tau$ directions with equal probabilities [Fig.~\ref{fig:method}(c)]. This hopping process can be made rejection-free by setting the arbitrary weight for a fictitious vertex as $\tilde{\lambda}_A = (1 - p)(z_s - 1)^{-1} \lambda_A$ where $z_s = 4$ ($6$) in 2D (3D).
After creating a fictitious vertex and the subsequent random-walk in the space-time~\footnote{We limit the maximum number of fictitious vertices to $\sim O(1)$ during the simulation.}, the worm may return to this fictitious vertex at a site yet to be marked by additional $\sigma^x$. In 2D, where $A_s$ is a four-spin operator, we can bring this vertex back to normal after the worm hops to the last unmarked site. In 3D, after a similar process we still need to wait for the worm to return once again to recover a normal $\lambda_A$ vertex. With such sequences we can generate off-diagonal multispin processes like the one in Fig.~\ref{fig:method}(b) or even more complicated ones. The observables are estimated when there is no fictitious vertex, namely, while we are sampling physical world-line configurations.

\textit{Conclusions.}
We presented thermodynamic phase diagrams of the 2D and 3D toric codes
extended by a nearest-neighbor FM Ising coupling, which introduces quantum fluctuations in the magnetic flux excitations of the $Z_2$ QSLs, and eventually stabilizes the FM phase.
In 3D, we found a first-order transition between the $Z_2$ QSL and FM phases, with strong \textit{proximity effects} to QSL in the phase competing region. The finite-$T$ transition from paramagnet to QSL is in the inverted Ising universality class, as in the case without the Ising coupling. These characteristics will provide a clue for experimental studies on the 3D $Z_2$ QSL candidates.
In 2D, the $Z_2$ QSL state is unstable at finite $T$ and forms a QCP between the competing FM phase. We found the quantum critical behavior in the (2+1)D Ising universality class with an unconventional exponent for the magnetic susceptibility. The distinct behaviors between 2D and 3D are essentially due to whether flux excitations are deconfined particles (in 2D) or confined loops (in 3D) at low $T$.

\begin{acknowledgments}
\textit{Acknowledgments.}
We thank M.\;Udagawa for fruitful discussions.
We also thank K.~Harada for his efficient Bayesian FSS analysis tool and A.~Furusaki for a critical reading of our manuscript.
Y.\;K.\;acknowledges financial support from the \grant{RIKEN iTHES project}.
This work was supported by \grant{Grants-in-Aid for Scientific Research} under Grant No.\,26800199, the \grant{Strategic Programs for Innovative Research (SPIRE)}, MEXT, and the \grant{Computational Materials Science Initiative (CMSI)}, Japan. Numerical calculations were conducted on the RIKEN Integrated Cluster of Clusters (RICC).
\end{acknowledgments}

\nocite{apsrev41Control}
\bibliographystyle{my-apsrev4-1}
\bibliography{my-refcontrol,references}

\begin{thebibliography}{50}
\bibitem [{tbp()}]{tbp}%
  \BibitemOpen
  \href@noop {} {}\bibinfo {note} {The details will be published
  elsewhere.}\BibitemShut {Stop}%
\end{thebibliography}
\end{document}